\documentclass{PoS}

\title{Searching for the Culprit of Anomalous Microwave Emission: An AKARI PAHrange Analysis of Probable Electric Dipole Emitting Regions}

\ShortTitle{AKARI Analysis of AME Regions}

\author{\speaker{Aaron C. Bell}$^a$, Takashi Onaka$^a$, Itsuki Sakon$^a$, Yasuo Doi$^a$, Daisuke Ishihara$^b$, Hidehiro Kaneda$^b$, Martin Giard$^d$, Ho-Gyu Lee$^e$, Ryou Ohsawa$^a$ and Tamami I. Mori$^a$, Mark Hammonds$^a$\\
\llap{$^a$}Department of Astronomy, Graduate School of Science, The University of Tokyo, Tokyo 113-0033, Japan\\
\llap{$^b$}Graduate School of Science, Nagoya University, Nagoya 464-8602, Japan\\
\llap{$^c$}ISAS/JAXA, 252-5210 Kanagawa, Japan\\
\llap{$^d$}IRAP Universite de Toulouse \& CNRS, 9, Av. Du Colonel Roche BP 44346 F-31028 Toulouse, Cedex 4, France\\
\llap{$^e$}Korea Astronomy and Space Science Institute 776, Daedeokdae-ro, Yuseong-gu, Daejeon, Republic of Korea 305-348\\
        E-mail: \email{abell@astron.s.u-tokyo.ac.jp}}

\abstract{In the march forward of interstellar medium inquiry, many new species of interstellar dust have been modelled and discovered. The modes by which these species interact and evolve are beginning to be understood, but in recent years a peculiar new feature has appeared in microwave surveys. Anomalous microwave emission (AME), appearing between 10 and 90~GHz, has been correlated with thermal dust emission, leading to the popular suggestion that this anomaly is electric dipole emission from spinning dust \cite{draine98a}. The observed frequencies suggest that spinning grains should be on the order of 1~nm in size, hinting at poly-cyclic aromatic hydrocarbon molecules. We present data from AKARI/Infrared Camera \cite{onaka07}, due to the effective PAH/Unidentified Infrared Band (UIR) coverage of its 9~$\mu$m survey to investigate their role within a few regions showing strong AME in the Planck low frequency data. We include the well studied Perseus and $\rho$Ophiuchi clouds . We use the IRAS/IRIS 100~$\mu$m data to account for the overall dust temperature. We present our results as abundance maps for dust emitting around 9~$\mu$m, and 100~$\mu$m. Part of the AME in these regions may actually be attributed to thermal dust emission, or the star forming nature of these targets is masking the vibrational modes of PAHs which should be present there, suggesting further investigation for various galactic environments.}

\FullConference{The Life Cycle of Dust in the Universe: Observations, Theory, and Laboratory Experiments - LCDU 2013,\\
		18-22 November 2013\\
		Taipei, Taiwan}

\begin{document}

\section{Introduction to Anomalous Microwave Emission and Spinning Dust}
Anomalous microwave emission (AME) is defined for this work as follows, galactic microwave foreground emission which cannot be explained by synchrotron, free-free, or vibrational dust emission. Spinning dust may be another piece of the ``ISM Interaction Puzzle''. Small spinning dust grains have been suggested to carry the AME, or ``Foreground X'' \cite{costa04}. Grains with a localized charge which are excited rotationally should emit electric dipole radiation \cite{draine98a}. Assuming the peak AME frequency around 30~GHz, a dust temperature of 100~K, and a simple spherical moment of inertia, implies a dipole cross section on the order of 1~nm. PAHs are suspected because their typical size matches this range, however more evidence is desired in order to confirm or deny their possible role as an AME carrier. Two famous regions of significant AME emission were recently identified in the Planck Low Frequency Data \cite{planck11}.

\section{Method of Investigation}
The primary objective for this study was the comparison of data from the AKARI Infrared Camera (IRC) 9 and 18~$\mu$m surveys, representing vibrational emission by the UIR bands (likely carried by PAHs), to the Planck Low Frequency Instrument (LFI) 30~GHz data \cite{planck13}, representing the suspected spinning grain emission. The AKARI mid-infrared all-sky surveys are a very powerful tool for studying UIR band emission / PAH features. The resolution of the AKARI/IRC allows for a more detailed study of the spatial distribution of the PAH features than is possible using WISE or IRAS. All-sky coverage allows for complete inspection of any region of interest, and will allow this research to be carried out for new regions well outside of the area included in the Spitzer observations. The 9 and 18~$\mu$m surveys offer effective coverage of the UIR/PAH features. However it should be noted that directly comparing the fine-structures in the AKARI images to the Planck LFI data is impossible, since the LFI spatial resolution is about 30~arcminutes, compared to the AKARI/IRC's 9.4~arcseconds.

\subsection{Additional Data}
To support the AKARI data, IRAS 100~$\mu$m data was used to control for the overall dust abundance. Temperature maps from the IRAS/IRIS revision were used to derive approximate interstellar radiation field maps for the regions of interest, using a very simple $\beta{} = 2$ assumption for the emissivity. Corresponding images from the AKARI 9 and 18~$\mu$m surveys were then divided by the IRF map, and compared alongside the Planck 30~GHz images. All images were created from the same region of the sky, in 5~degree square maps, for both of the targets described below.

\section{Results and Discussion}
For this early study, two regions shown to be dominated by anomalous emission by the Planck Collaboration \cite{planck11} were selected. Choosing these regions simplified the process by reducing the contamination from other microwave foregrounds such as synchrotron emission. Their respective high galactic latitudes also reduce confusion with foreground and background ISM.

\subsection{$\rho$Ophiuchi}
The investigation included the star-forming region $\rho{}$Ophiuchi, centered at {\it l} = 353.05$^\mathrm{o}$ and {\it b} =  16.90$^\mathrm{o}$. This region shows two strong peaks in 1~cm (30~GHz) emission by Planck LFI.
\begin{figure*}[htbp]
\begin{center}
\includegraphics[scale=0.80]{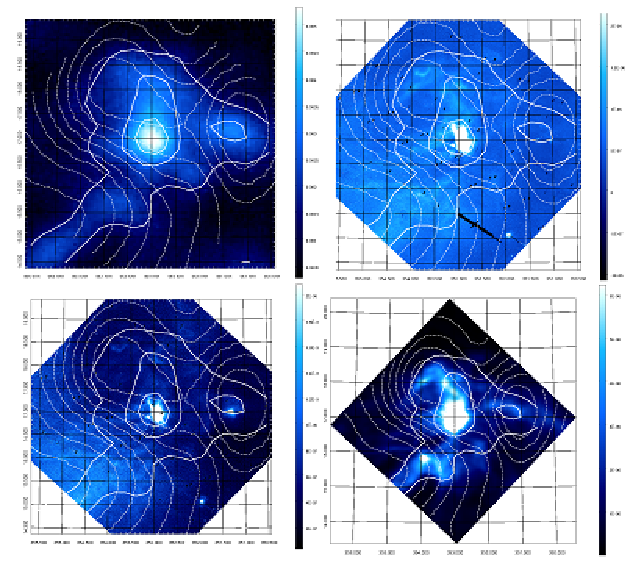}
\caption{$\rho$Ophiuchi at multiple wavelengths. Clockwise: Planck 30~GHz, AKARI/ 9~$\mu$m, AKARI 18~$\mu$m, and IRAS 100~$\mu$m. The ISRF is traced by dashed contours in each image. The IRAS and AKARI images shown here have been divided by the ISRF. The solid contours follow the Planck 30~GHz data. }
\end{center}
\end{figure*}

The strong 30~GHz peak around the center of the $\rho$Ophiuchi star-forming region appears bright in each image. PAH abundance may be high here but, not necessarily enhanced with respect to overall dust abundance. It has been suggested that this is a very complicated region, and not necessarily a good target for AME investigation, despite being dominated by AME. Note that the right-side 30~GHz peak agrees with the ISRF peak, suggesting that this peak may be enhanced thermal dust emission rather than AME. Multi-wavelength SED modeling will investigate this further.

\subsection{Perseus Molecular Cloud Complex}
A 5 degree square region, centered at {\it l} = 160.26$^\mathrm{o}$ and {\it b} = -18.62$^\mathrm{o}$, was chosen and investigated across multiple wavelengths (AKARI/IRC 9 and 18~$\mu$m, IRAS/IRIS 100~$\mu$m, Planck/LFI 1~cm).
\begin{figure*}[htbp]
\begin{center}
\includegraphics[scale=0.80]{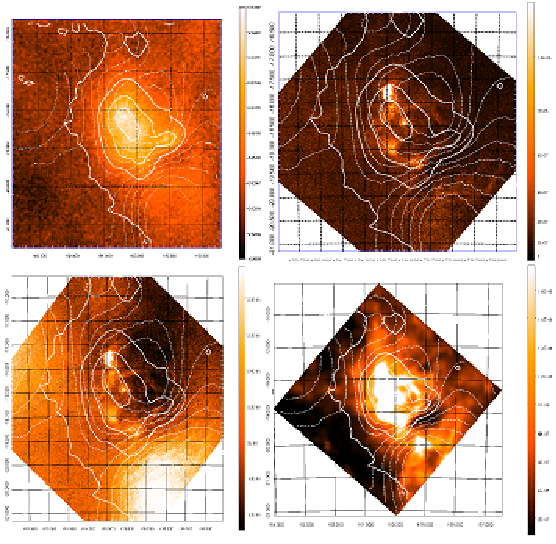}
\caption{Perseus Molecular Cloud at multiple wavelengths. Clockwise: Planck 30~GHz, AKARI/ 9~$\mu$m, AKARI 18~$\mu$m, and IRAS 100~$\mu$m. The IRF is traced by dashed contours in each image. The IRAS and AKARI images shown here have been divided by the IRF. The solid contours follow the Planck 30~GHz data. }
\end{center}
\end{figure*}

This part of the sky contains one bright peak in the 30~GHz image, surrounding the IC348 star-forming region. In the Perseus Cloud, the 30~GHz emission generally correlated with emission at all other wavelengths. There may be a slight offset between the position of the peak 9~$\mu$m emission, and the peak 30~GHz emission. However the beam of the Planck data (33$'$), is much wider than that of AKARI/IRC (9.4$''$) and spatial comparison of the images is difficult. The 100~$\mu$m IRAS data appears to more closely follow the shape seen in the 30~GHz data. Further investigation will explore various other galactic environments. The AME may be dominant in these regions, but their complex nature as star-forming regions may confuse situation. Perhaps only a second order relationship between PAH abundance and AME may be noticeable. Future research will explore these possibilities and report findings in more detail, incorporating data at longer wavelengths via the AKARI/Far Infrared Surveyor all-sky maps.

\section*{Acknowledgements}
This investigation is primarily based on  all-sky surveys by AKARI, a collaboration between JAXA and ESA. An astronomically large quantity of gratitude is extended to T. Onaka and members of the AKARI team for guidance, and to the Japanese MEXT for scholarship support making A.Bell's studies possible, and the U. Tokyo Grad. School of Science. Additional thanks to Steven Gibson for continued encouraging comments.

\end{document}